# DCaaS: Data Consistency as a Service for Managing Data Uncertainty on the Clouds

Islam Elgedawy

Computer Engineering Department,
Middle East Technical University,
Northern Cyprus Campus,
Guzelyurt, Mersin 10, Turkey.

*Abstract*—Ensuring data correctness over partitioned distributed database systems is a classical problem. Classical solutions proposed to solve this problem are mainly adopting locking or blocking techniques. These techniques are not suitable for cloud environments as they produce terrible response times; due to the long latency and faultiness of wide area network connections among cloud datacenters. One way to improve performance is to restrict access of users-bases to specific datacenters and avoid data sharing between datacenters. However, conflicts might appear when data is replicated between datacenters; nevertheless change propagation timeliness is not guaranteed. Such problems created data uncertainty on cloud environments. Managing data uncertainty is one of the main obstacles for supporting global distributed transactions on the clouds. To overcome this problem, this paper proposes an quota-based approach for managing data uncertainty on the clouds that guarantees global data correctness without global locking or blocking. To decouple service developers from the hassles of managing data uncertainty, we propose to use a new platform service (i.e. Data Consistency as a Service (DCaaS)) to encapsulate the proposed approach. DCaaS service also ensures SaaS services cloud portability, as it works as a cloud adapter between SaaS service instances. Experiments show that proposed approach realized by the DCaaS service provides much better response time when compared with classical locking and blocking techniques.

*Keywords*—*clouds; cloudlet; cloud adapter; data uncertainty; DCaaS; SaaS; PaaS*

## I. INTRODUCTION

Clouds are the next-generation datacenters virtualized through hypervisor technologies, where cloud-vendors can dynamically provision their virtualized nodes on demand to their customers according to the specified service level agreements [3]. Cloud computing is the computing paradigm that enables the whole solution stack (from hardware to software) to be delivered as services over the internet. Such services are classified into three basic classes: Software as a service (SaaS), Platform as a Service (PaaS), and Infrastructure as a Service (IaaS) [3]. *SaaS services* are applications that customers need. *PaaS services* are services needed to deploy and deliver SaaS services such as database and middleware services. *IaaS services* are services needed to specify the required virtualized computer infrastructure such as disk and memory requirements.

Real-life cloud environments usually constituted from a collection of datacenters connected via a Wide Area Network (WAN). A datacenter is constituted from thousands of machines connected via a LAN (i.e. local area network) forming what is known as a *cloudlet* (i.e. a small cloud). Latency in WANs is much bigger than latency in LANs. This difference in latency distribution inside cloud environments created a non-homogenous timing model for the cloud. For example, latency between two machines inside a datacenter is in the range of 100 msec; while latency between two machines connected via WAN is in the range of 1000 msec (i.e. when machines are in different continents). Such latency difference makes the WAN connections as the main bottleneck in cloud environments. Hence, existing classical concurrency control and transaction management approaches (such as ones discussed in [7] [9]) are not suitable for cloud environments, as they opt to accommodate the slowest latency inside the cloud environment, which badly hurts services performance. To overcome this problem, many approaches have appeared [4][5][6][8][20][22][24] proposing a restricted version of cloud computing, in which requests of users with similar latency values (known as a user-base) are directed to the closest datacenter such that no data sharing between datacenters is allowed. However, data could be replicated later between datacenters storages as a background process to keep databases eventually synchronized and to create multiple copies of the database for backup purposes [12]. We define such computing model as "cloudlet computing". A cloudlet is a small cloud, so it is similar to cloud in terms of offered services; however it differs in restricting its physical scope into only one datacenter. Cloudlet computing only supports what is known as a mini-transactions [20], which are transactions restricted to a single datacenter to guarantee good performance [4][5][6][8][20][24]. On the other hand, cloudlet computing cannot support global transactions (such as in flight reservation and banking), as global data correctness is not guaranteed due to lack of global control. In other words, cloudlet computing paradigm ensures local correctness of the data within the cloudlet but cannot ensure global correctness of the data (among all cloudlets), as conflicts might appear when data is replicated between cloudlets due to lack of global control. Furthermore, there is no guarantee for change propagation timeliness as updates propagation depends on many different factors such WAN latency and replication schedule of cloud vendor. Hence, data uncertainty becomes a very important characteristic on the clouds and must be managed by applications as cloud data stores management systems cannot overcome such problem [13]. However, managing data uncertainty in service code is not an easy task, as it requires services to be designed in a different way to deal with missing or multiple values of data objects. Actually, managing data uncertainty is one of the main obstacles for writing transactional applications on the clouds [3]. We argue that service developers should be totally decoupled from managing data uncertainty in their code to improve service maintainability, portability and reusability, nevertheless reducing service development efforts and time. We argue that we need to create a new breed of database management services (such as transaction management, data access, and data replication) that take into consideration the data uncertainty resulting from the cloud non-homogenous timing model. Unlike the classical centralized database management systems, such new breed of database management services must be totally decoupled and must ensure good performance, high availability, and high scalability of services as well as global data correctness, which enables us to easily support global distributed transactions. This paper proposes our first initiative towards achieving these goals. Hence we summarize the contributions of the paper as follows:

*1) First, we propose to use a new middleware platform service for handling data consistency and data uncertainty issues (i.e. Data Consistency as a Service (DCaaS)) on behalf of service developers, hence service code will be totally decoupled from data uncertainty management code leading to faster maintainable SaaS service development. SaaS developers will not write SQL statements in their SaaS service code to access data; instead they will write invocations for the DCaaS service APIs operations to access their data. Furthermore, DCaaS service ensures service cloud portability, as it also decouples SaaS services from directly accessing PaaS services operations; hence no SaaS service code will change if the cloud vendor is changed. The only required change is in the interface between the DCaaS and the PaaS services, which could be handled easily using service adapters [28].*

*2) Second, we propose to use a multi-level data consistency approach for handling SaaS services data objects to enhance service performance. As maintaining strong data consistency is a costly process [15], we argue that it should be only used for objects that their correctness is crucial for services correctness, while for less important data we could go for weaker consistency notions such as eventual or session consistency [23]. Service developers will dynamically define their consistency requirements according to their business logic in a form of a Data Consistency Plan (DCP), and then submit such plan to the DCaaS service, which will make sure such consistency requirements are fulfilled during data access operations. Currently, we support three levels of data consistency strong, eventual, and session that service providers choose from to define the required DCP; more details are given in Section 4.*

*3) Third, we propose a quota-based approach for ensuring global data correctness among cloudlets. The proposed approach applies inventory management principles to ensure fulfillment of users requests, that it requires service providers to divide crucial objects capacity among cloudlets by specifying a quota for each cloudlet such that DCaaS services makes sure no cloudlet user request consume more than the allocated quota. Hence, when data is replicated between cloudlets no conflicts could arise. When a given DCaaS service instance requires more than assigned quota due to high volume of requests, it could contact other DCaaS instances to borrow extra quota. If quota borrowing process fails the request is rejected. To achieve such goals, we provide different protocols for quota borrowing, object stabilization, and DCaaS fault tolerance to ensure protocols liveness and safety properties, more details are given in Sections 4, 5 and 6.*

Experiments show that proposed DCaaS service adopting the proposed data consistency approach provides much better response time when compared with classical locking and blocking techniques. The rest of the paper is organized as follows. Section 2 provides a brief background and discusses related work. Section 3 provides solution model and assumptions. Section 4 introduces the quota-based approach proposed for ensuring data global correctness. Section 5 discusses different management issues of data consistency plan and proposes the adopted object stabilization protocol. Section 6 discusses various design aspects of the proposed DCaaS service such as required APIs and DCaaS recovery. Section 7 provides some basic comparative simulation experiments for proposed approaches, and finally Section 8 concludes the paper. This paper is the extended version of the paper proposed in [27].

## II. BACKGROUND AND RELATED WORK

Ensuring data consistency over partitioned distributed database system is a classical problem that attracted many researchers. Data replication is one of the methods used for sharing data between database instances; in which multiple copies of the shared data are stored with the SaaS service instances [7][9][12]. Such copies (replicas) are frequently updated by broadcasting changes to all instances. However, this is not an easy step, as correctness of the data must be maintained. One important aspect of replicated data correctness is mutual consistency, in which all copies of the same logical data must agree on exactly one current value for the data items without violating the logic of the executed transaction. Furthermore, the problem becomes more complicated, when a failure occurs (e.g. due to network failure or server failure) as the correctness of the shared replicated data could be compromised via uncoordinated updates. Classical solutions proposed to solve this problem are mainly adopting locking or blocking techniques to ensure data correctness. Good surveys for such approaches could be found in [7] [9]. Such classical approaches adopt a pessimistic strategy that assumes conflicts occur frequently. Hence, they suspend all other instances from working (via locking or blocking) when a given instance needs to do some updates for the shared data. These techniques provide very bad performance when applied on cloud environments [5] [6] [11] [15] [22], as they tend to create considerably high overhead over the slow faulty WAN connections due to exchanged synchronization messages and performed reconciliation transactions, which of course badly hurts services availability and customers' response times. CAP theorem [11] clearly states that there is a tradeoff between Data consistency, service availability and partition tolerance. This means asking for high availability and consistent data would imply that we cannot tolerate network partitioning. In cloud environments, data is partitioned over multiple machines to provide high scalability, and as networks between these machines could simply fail, this means partitioning (data and networks) is crucial for cloud environments. Hence, cloud vendors opt to choose between availability and consistency. Recent approaches (such as Google's BigTable [4], Yahoo PUNTS[5], Amazon's Dynamo [8], G-store [6] and Apache Cassandra [24]) proposed to go for weaker forms of consistency on the clouds such as eventual consistency [20], in which they trade consistency for availability, that all service instances are allowed to work normally without any suspension and process their transactions locally. This is known as the optimistic strategy; as it assumes conflict occur rarely. However, when a conflict is detected undo transactions and/or compensating transactions should be performed by the services, also in some cases some data versions could be lost. Going for weaker forms of consistency requires developers to design programs in new ways that can tolerate such data inconsistencies according to business logic. This could be done via writing correcting transactions or using data time stamps to decide between multiple versions of the data as in [21]. Solution proposed in this paper compromise between the optimistic and pessimistic approaches such that it maintains local correctness within a cloudlet using pessimistic approaches and ensures data global correctness between cloudlets by using a quota-based approach that adopts lazy replication approaches as in optimistic approaches to ensure availability and scalability.

### III. SOLUTION MODEL AND ASSUMPTIONS

In this paper, we assume cloud vendors provide a PaaS service for accessing the SaaS database (that is a tenant in the physical cloud database). Objects of the SaaS database are stored as simple key-value data format. SaaS database could be partitioned among different cloudlets; hence we require cloudlets PaaS services to provide a lazy replication mechanism (as a background process) to replicate their data changes. A PaaS service could be accessed by one or more DCaaS services simultaneously; hence we require a PaaS service to provide a local concurrency protocol mechanism between DCaaS instances accessing it. Each SaaS instance handles a given user-base of SaaS customers. Each cloudlet could create multiple SaaS, DCaaS, and PaaS service instances to increase availability, throughput, and enhance response times, as depicted in Figure 1. Hence we require each DCaaS instance to give reference to other DCaaS instances created inside and outside its cloudlet. We model DCaaS service instances as peers and they can communicate with each other in a P2P manner. We require all the communications between SaaS, DCaaS, PaaS services to be done in an asynchronous mode, as the clouds timing model is non-homogeneous. Hence, fast services will not wait for slow services responses and could process other requests. We require a state machine to implementation be installed at each DCaaS instance to realize proposed protocols, the exchanged messages between state-machines are calls for DCaaS API operations.

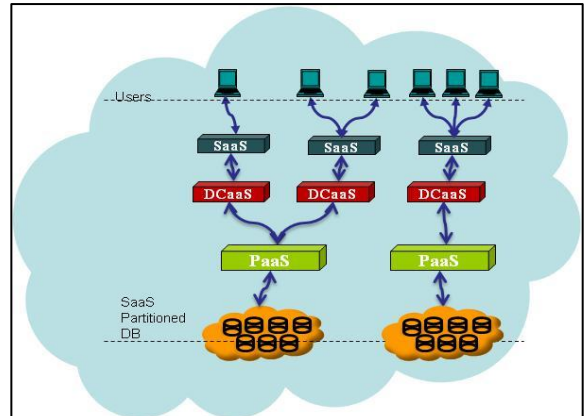

Fig. 1.    aS, DCaaS, PaaS single cloudlet deployment

### IV. QOUTA-BASED DATA CONSISTENCY APPROACH

Work in [13] clearly indicates that data uncertainty must be managed in distributed transactions in order to meet real life requirements. Management of data uncertainty is not a new problem. Actually, in business, handling data uncertainty is a fact of life and many solutions have been adopted by businesses for managing such uncertainty such as reserved inventory, allocations against credit lines, and budgeting. We propose to handle uncertainty for data objects using similar business strategies. For example, inventory management is primarily about specifying the shape and percentage of stocked goods required at different locations within a facility or within many locations of a supply network. Inventory management is the process of efficiently overseeing the constant flow of units into and out of an existing inventory. This process usually involves controlling the transfer in of units in order to prevent the inventory from becoming too high, or dwindling to levels that could put the operation of the company into jeopardy. Hence, we argue we could use the same process for managing transactions accessing objects on distributed data stores such that the data store act as inventories, the objects act as the goods, and the users' requests act as the consuming demand. For example, an airline reservation service could have its database partitioned among many cloudlets (i.e. inventories). Instead of globally locking flight data object whenever a booking operation is made, we will allocate a quota of seats (i.e. goods) for each cloudlet such that each cloudlet locally handles its incoming users requests (i.e. demand) and its DCaaS service instances make sure it does not exceed the allocated quota. When such condition is fulfilled, no data conflicts (i.e. different bookings for the same seat) could appear when replication occurs between cloudlets. A cloudlet ensures the correctness of its transactions using its own concurrency control approach using any locking or blocking technique. This approach will not hurt performance as latency inside cloudlets is small (i.e. within the range of 100ms), which still provide acceptable response time [20]. Another example, in banking, if we need to access a given account, instead of locking the account object, we could allocate a budget for each cloudlet to manage its incoming withdrawal and deposit requests.

To ensure global data consistency, we require each SaaS service provider to define a capacity quota for its strong consistency data objects for each cloudlet; then provides such information to the corresponding DCaaS service instances via a DCP. DCaaS makes sure that none of incoming users requests to consume more than the allocated objects quotas. In case of one request requires capacity more than allocated quota, the involved DCaaS service instance tries to borrow quota from other DCaaS instances. In case of success, it accepts the request and processes it, otherwise it rejects it. A DCaaS service instance could borrow from instances located in its cloudlet, or from instances in other cloudlets. As borrowing from outside cloudlets requires communications via WAN connection, hence only requests requiring extra quota will be affected. We argue that quota should be distributed between cloudlets in a manner that minimizes the borrowing rate, that quotas should be proportional to the volume of the cloudlets users-bases such that cloudlet with bigger volume should take a bigger quota. We perform the quota borrowing process adopting a simple protocol depicted in Figure 2.

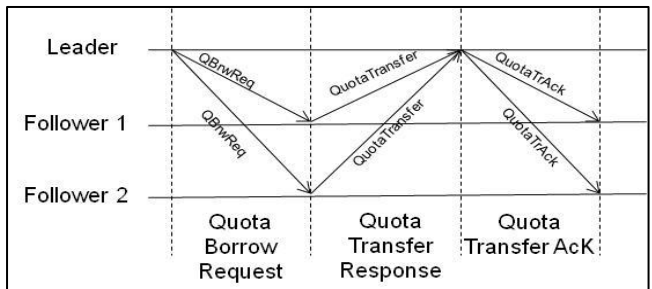

Fig. 2.    Stages of Quota Borrowing Protocol

The quota borrowing protocol works as follows. The DCaaS service instance requesting the quota sends its request first to DCaaS instances in its cloudlet with the required of borrow amount. Each DCaaS service instance received the quota borrow request replies back with the quota amount it can transfer. This amount ranges from zero to the required amount. The leader collects all quota transfer responses and acknowledges other DCaaS instances with the amounts it will take. Once DCaaS instances receive such acknowledgment it updates its share of quota with the acknowledged amount. Such protocol requires a state machine to be installed at each DCaaS instance, exchanged messages between state-machines are calls for DCaaS API operations; more details about DCaaS APIs will be given in Section 6. The easiest quota distribution strategy is to equally divide the object capacity among cloudlets. However, the proper quota distribution strategy should be based on thorough demand forecast analysis. In case a service provider makes a mistake in allocating the quotas, DCaaS service instances will automatically redistribute the quotas among themselves when requests arrives via the quota borrowing process. The price of wrong quota allocation is longer response times due to the slow quota borrowing process (if WAN connections are used). However, once quota borrowing process is finished, response times dramatically improve, as all incoming requests will be handled locally inside the cloudlet, as shown in Section 7.

## V.  SERVICE DCP MANAGEMENT

Our solution divides the responsibility of managing data uncertainty between the service provider and the DCaaS service. It decouples the definition of the data management strategy from its implementation. The proposed solution requires SaaS service providers to specify the required strategy, while the DCaaS service work on implementing and executing this strategy. A SaaS service provider specifies its strategy by defining a Data Consistency Plan (DCP) for its SaaS service then submits such DCP to the DCaaS service to implement it. DCP specifies the required consistency level for each data object and its corresponding object stabilization method. Service providers could change their DCP at run time without changing their service code. For each data access, DCaaS service checks the required consistency level defined in the DCP; then invokes the corresponding data access procedure. This section discusses different management aspects of SaaS data consistency plan. First it introduces the supported data consistency levels and provides a formal definition for a DCP. Then it describes DCP change management process. Finally, it shows the adopted object stabilization protocol as well as the supported stabilization methods in case of conflicts.

### A.  DCP Definition and Creation

Currently, we support three levels of data consistency (i.e. Strong, Eventual, and Session). Strong consistency implies that the global correctness of the data object is maintained such that any SaaS instance accessing the object is actually reading its up-to-date correct value. Eventual consistency implies that object correctness is locally maintained (i.e. within a cloudlet) but not globally (i.e. between all cloudlets). However, if there are no global conflicts between cloudlets, and no more new updates are made to the object, eventually all database accesses will return the same last updated value, as cloud vendors perform a lazy replication process between cloudlets to synchronize their DBs [12]. Session consistency implies that the SaaS instances read its own writes only. This means object data will be maintained only at the DCaaS service instance cache and does not go to the PaaS service for storage. Hence, those data will be lost after the session terminates. We require each SaaS provider to define a DCP for its service; hence each SaaS service instance will follow the same service DCP. A DCP indicates the required consistency level for each data object. Also it indicates the required stabilization method to be applied in case of object values divergence. DCP also specifies the cloudlet quota for strong consistency objects. We formally define a DCP as a set of Object Access Patterns (OAP) that $DCP = \{OAP(i)\}$, where an $OAP(i) = <i, c, s, q>$, $i$ is the data object reference, $c$ is the required consistency level, $s$ is the required stabilization method, and $q$ is the cloudlet quota distribution plan and it is defined as a set of cloudlet quota allocations, that $q = \{<Cloudlet\ reference,\ CloudletQuota>\}$. As the number of cloudlets is always small, the size of such quota list is not a problem. We support different stabilization methods, more details about object stabilization will be provided later. We require each data object to have only one OAP. For example, a SaaS provider for a flight reservation service $X$, which require access for two data objects *Customer* and *Flight*, The corresponding DCP could be defined as $DCP[X] = \{<Customer, Eventual, Thomas, \{\}>, <Flight, Strong, Exact, \{<1,50>, <2, 200>\}>\}$. This means the consistency of the customer object is eventual and Thomas write rule (i.e. last write wins) will be applied in case of conflicts, while the consistency of the flight object is strong, and history method will applied in case of conflicts. It also shows that we have two cloudlets, the first cloudlet has a quota of 50, and the second one has a quota of 200. As we can see the given DCP definition is working at the object level, however we could extend the definition to work on the attribute level, by defining DCP as a set of Attributes Access Patterns (AAP) that $DCP = \{AAP[i , j]\}$ , where an $AAP[i,j] = < i, j, c, s, q>$, $i$ is the data object reference, $j$ is the attribute reference, $c$ is the required consistency level, $s$ is the required stabilization method, and $q$ is a set of allocated cloudlets quota. For example, a DCP over flight attributes could be defined as $\{<Flight, PlaneModel, Eventual, Thomas. \{\}>, <Flight, Capacity, Strong, Max, \{<1, 50>, <2, 200>\}>\}$. We do not require specific granularity level for the DCP definition; we leave this choice to SaaS providers to decide. If the SaaS providers choose an object level or a higher level, DCP size could be small and fits nicely in memory but performance could be affected due to local concurrency control locks. However, if they choose the attribute level, the DCP size could be big; hence DCP could not fit into memory and require storage. Of course, this is a classical optimization problem a SaaS provider has to solve. Once quick solution is to compress DCPs using any query-aware compression technique (such as one in [26]) to avoid DCP storage. Another approach to minimize the DCP size is to assume default values for unspecified objects and attributes. We use eventual consistency as the default consistency level, and Thomas rule as the default stabilization method. It is important to note that in this paper, we require DCP to have only one access pattern for each object/attribute. However, in future work we are planning to relax this condition to allow a given object to have different access patterns that DCaaS could choose from in a context-based manner (i.e. choice could be based on the executed SaaS operation, PaaS response time, Users SLAs).

### B. DCP Change Management

To provide flexibility for SaaS providers, we provide them with the option to change their DCPs at run time whenever they like and the DCaaS service will do the necessary adjustments to fulfill the new requirements. The DCaaS service contains different components to handle different consistency requirements (refer to Figure 4). It is important to note that change in DCP does not require change in the SaaS service data access code, as DCP change occurs through a specific DCP APIs, while the data access occurs via invocations for different API operations, more details about DCaaS APIs will be given in section 6.

As we do not allow different access patterns for the same data object, whenever DCaaS service instance receives a request for DCP change, it automatically becomes the DCaaS instances leader and notifies the other DCaaS service instances with the DCP change and make sure it is executed at all instances. For consistency level upgrade request from session to eventual, the DCaaS instance leader updates the corresponding DCP entry, then stores the object value written in its cache into the data store via the PaaS service, and then notifies other instances and waits for their acknowledgments. If all instances replied, it considers the request is fulfilled. In case of missing or slow acknowledgment, the leader tries back after certain timeout window, if an instance still not replying, the leader consider it as a failed node and store the change request for later when it recovers, more details about DCaaS recovery will be given later. For consistency level upgrade request from session/eventual to strong, the DCaaS instance leader updates the corresponding DCP entry, and then starts to stabilize the object values in all cloudlets as correctness of such values were not maintained before the upgrade request. This is done by broadcasting a stabilization request for all DCaaS instances. We have different strategies for stabilizing different object values that differ in their costs, more details are given later in Section 5.3. Once the leader stabilizes the object value, it computes the DCaaS instances quotas then sends for each DCaaS instance the new object value and its allocated quota, more details about quota computation are given in Section 6. For consistency level downgrade request from strong to eventual, the DCaaS instance leader updates the corresponding DCP entry to stop quota checks, as now only local correctness is required. For consistency level downgrade request strong/ eventual to session, the DCaaS instance leader updates the corresponding DCP entry, and then creates an entry in its cache for the object and stop storing object updates into the data store as all updates has to in the cache only. In both cases, DCaaS leader notifies other DCaaS instances with the change and waits for their acknowledgement.

### C. Objects Stabilization

When a given DCaaS instance receives a request for consistency upgrade to strong consistency, it requires stabilizing the object value, as every DCaaS instance could have a different value. Our stabilization protocol is very simple. First, we assign the DCaaS instance receiving the change request as the leader who will orchestrate the change. Other DCaaS instances will be the followers. The leader sends a stabilization request messages to all DCaaS instances and waits for their response.

Each DCaaS instance must reply back to the leader with the current value of the object using a stabilization response message. The leader collects all values and computes the new object value by applying the stabilization method defined in the DCP. Finally, the leader sends to each DCaaS instance a stabilization command message to propagate the computed value and monitor instances acknowledgments. Once a DCaaS instance receives a stabilization command, it updates the object value and its DCP and replies with an update acknowledgement. Of course, implementation of such protocol requires a state machine to be installed at each DCaaS instance, exchanged messages between state-machines are calls for DCaaS API operations; more details about DCaaS APIs will be given in Section 6. Figure 3 summarizes the steps of the stabilization protocol.

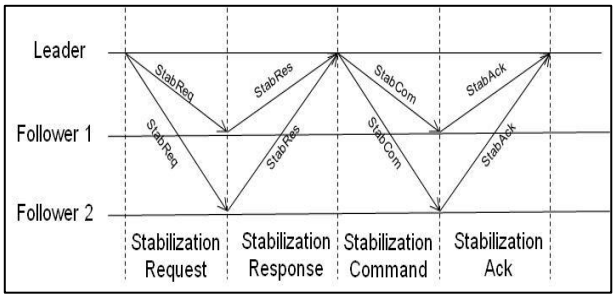

Fig. 3.    Stages of Object Stabilization Protocol

Figure 3 shows the different stages of the protocol. To simplify the diagram we assumed propagation delays between DCaaS instants are constant. However, in real life, propagation delays are different. Such stabilization process is a very costly process as it involves communication over WAN connections. Hence, we advise SaaS providers to avoid frequent consistency upgrades. We assume all exchanged messages are asynchronous.    We propose different types of stabilization methods in order to provide SaaS providers with the flexibility to choose the most suitable ones for their business logic. Each data object will have its own stabilization method defined in the DCP. To stabilize an object, we propose to use different methods for stabilization varying in complexity, cost, and correctness. 1) Exact method. 2) Thomas write rule. 3) Basic uncertainty filters: Min, Max, Avg, and Sum. 4) Customized uncertainty filter. The exact method guarantees the correctness of the data object value. However, it is the most expensive method and we do not recommend it for SaaS providers due to the huge amount of communication and computation involved. The exact method requires keeping track of transaction history at each DCaaS instances, then sending these histories to the leader to find a global order for all transactions. Then the leader has to execute these transactions to compute the new value, and then distribute the new history and new value to other DCaaS instances. Of course, finding such global transaction order is a very expensive task as it could require many transactions rollbacks over all DCaaS instances. Hence, SaaS providers should use this method only for objects that is extremely crucial for their business logic. The Thomas write rule is one of the most famous methods in conflict resolution. It simply returns the value with the most recent time stamp. Hence, each DCaaS instance should send the leader the object value with its corresponding time stamp. The leader simply chooses the most recent one. The problem with this approach if real time is used is to have global clock synchronization, which is not feasible. However, there are many solutions proposed in distributed computing area for this problem such as use of lamport clock [16]. To avoid the headache of global clock synchronization, we provide the option to use basic uncertainty filters that are used in the area of probabilistic databases [1] [2]. That DCaaS follower sends only the values, and the leader applies one of the basic probabilistic basic functions (such as Min, Max, Avg, Median, and Sum) to get the new object value. Finally, we provide the SaaS providers to provide their own customized uncertainty filters if they did not like to use basic ones.

## VI. DATA CONSISTENCY AS A SERVICE

DCaaS service is basically proposed to decouple SaaS developers from managing data uncertainty aspects in their services code. SaaS developers will not write SQL statements in their SaaS service code to access data; instead they will write invocations for the DCaaS service APIs operations to access their data. Also DCaaS service decouples SaaS developers from PaaS services, hence it ensures SaaS clouds portability as no changes will occur to the SaaS service code if we change cloud vendors, the only change will be in the DCaaS service interface with the PaaS service, which could be managed by service adapters [28].    DCaaS takes the consistency requirements of a given SaaS service as a DCP, and then automatically implements and executes the given DCP for each data access. SaaS developers have the flexibility to change their consistency requirements on run time without changing their SaaS service code. This section briefly discusses various design aspects of DCaaS service.  First, it discusses the DCaaS structure and configuration, and then it describes the different DCaaS APIs, and finally it illustrates adopted protocols for DCaaS service recovery.

### A.  DCaaS Structure and Configuration

As we support three different levels of data consistency, DCaaS service should have implementations for approaches realizing the adopted data consistency levels.  To decouple DCaaS service code from the realizing approaches implementations, we encapsulate each data consistency approach as a *component service* to be invoked by the DCaaS service. We can think of the DCaaS service as the orchestrator for these service components. For example, a developer could invoke a DCaaS write operation to update a value of an object (e.g. *DCaaS.Write (X,1)*. Listing 1 depicts a sketch for a DCaaS service write operation. DCaaS should invoke the write operation version corresponding to the required data consistency level.

Listing 1: A sketch for DCaaS Write operation

```
int Write  (DataObject X, ObjectValue V)
  {
      Consistency Level   L = GetConsistencyLevel(X);
          Switch (L)
              {Case Strong :  status= Strong-Write (X, V);
                  Case Eventual: status= Eventual -Write (X, V);
                  Case Session: status= Session -Write (X, V);
                  //...
                  }
              return (Status);
  }
```

Each component service communicates with the PaaS service to perform the required operations on the data store, as in Figure 4. The DCaaS service is not necessary to be located on the same machine of its service components or the PaaS service. We require that each cloudlet to have at least one SaaS service instance, one DCaaS service instance and one PaaS service instance.

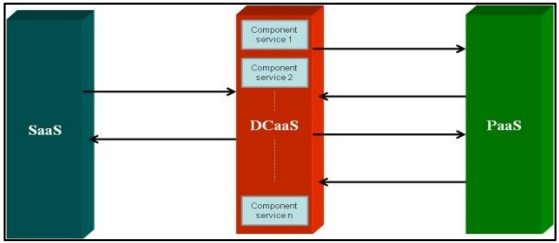

Fig. 4.        SaaS, DCaaS, PaaS interactions

A cloudlet management system could clone SaaS and DCaaS service instances to improve performance, however, we require a certain configuration protocol to be followed whenever a new DCaaS service instance is created in order to ensure DCaaS functional correctness. When a DCaaS service instance is created, it will not be in the active state unless the configuration process is finalized. This configuration process could be done manually or automatically. In the manual mode, after a SaaS provider creates a DCaaS instance at each cloudlet, it provides each DCaaS instance with the corresponding DCP as well as a list of other DCaaS service instances (i.e. we define it as the peer list). Once a DCP is loaded, the DCaaS service instance creates a list of Strong Consistency Objects Quotas (i.e. *SCOQ = {<Object Reference, InstanceObjectQuota>}*) to keep track of its quota. This is done by copying DCP strong object entries into the list and setting *InstanceObjectQuota* to the quota of its cloudlet (i.e. *CloudletQuota* defined in the DCP). In case a cloudlet has only one DCaaS service instance, *InstanceObjectQuota* will be equal to *CloudletQuota*. However, if the cloudlet has multiple DCaaS instances, The *CloudletQuota* should be equal to the sum of all *InstanceObjectQuota* belonging to its DCaaS service instances. In the automated mode, the SaaS providers provides only one DCaaS instance with the DCP and the peer list, and this DCaaS service automatically contacts the other DCaaS Service instances in the peer list to upload the required DCP and the given peer list by invoking specific DCaaS APIs. Again, once each DCaaS loads its DCP, it creates its local SCOQ list. Once a DCaaS instance has its DCP, peer list, and SCOQ list ready, it becomes now in the ready state. When a new DCaaS instance is added to the DCaaS peer network, the cloudlet quota of strong consistency objects specified in the DCP has to be redistributed among all DCaaS service instances inside this cloudlet, and then each DCaaS service instance should update its SCOQ list with the new quota values. This is done via a join DCaaS instance protocol, in which a new DCaaS service instance sends to the current cloudlet leader a join request. If there is no current leader, the new instance sends the join request to any existing DCaaS service instance, which will become the leader. We adopted this simple leader selection approach to avoid doing leader election process. Once a leader receives the join request and makes sure the new instance is authentic and not malicious (security aspects are out of the scope of this paper), it adds it to its peer list and updates its SCOQ list by dividing the cloudlet quota of each object by the number of instances in the new peer list, then sends add instance request to all DCaaS instance in its old peer list. Once a DCaaS instance receives the update peer list request, it adds the new node to its peer list and updates its SCOQ list by dividing the cloudlet quota of each object by the number of instances in the new peer list, and then acknowledges the leader. Once the leader receives all acknowledgments, it replies back to the new instance with the join accepted message and provides it with the peer list and the DCP, from which the new instance will compute its SCOQ

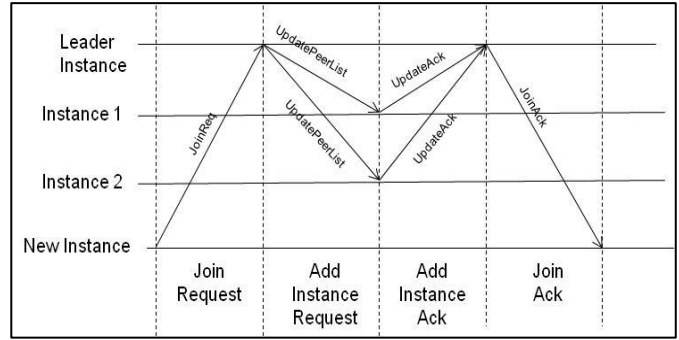

list. Figure 5 summarizes the steps of the join protocol.

Fig. 5.        Stages of DCaaS Instance Join Protocol

Once each DCaaS instance computes its new SCOQ list, it becomes in the ready state and could process user requests. It is important to note that DCaaS instances check first if the required instance to be added is not in their peer list before they do the SCOQ list computation, otherwise they keep the old SCOQ list, as no changes are occurred. This is important issue to make sure the join requests from the recovered instances or new instances are idempotent. DCaaS instance recovery is discussed in the later in this section.

### B. DCaaS APIs

DCaaS service API should support operations required for different service interactions. For example, it should support data access operations, operations for objects stabilization, operations for DCP management and quota redistribution, operations for peer list management, and operations for quota borrow and transfer. For data access APIs, DCaaS service implements a simple API interface for reading and writing operations. The read operation API is *Read (DataObject X)*, while the write operation API is *write (DataObject X, ObjectValue V)*. For managing data consistency plans, DCaaS service should provide DCP management APIs such as *LoadDCP(DCP p)*, *ModifyConsistencyLevel(DataObject X, ConsistencyLevel L)*, *ModifyCloudletQuota(DataObject X, CloudletReference R, Quota Q)*. *LoadDCP* is used to load a DCP when a DCaaS service is created, while *ModifyConsistencyLevel*, is used to modify a given data object consistency level. *ModifyCloudletQuota* is used to modify a given cloudlet quota. For peer list management, we should have APIs such as *LoadPeerList(List L)* to upload a peer list when DCaas is created, *UpdatePeerList(UpdateType T, UpdateDetails D)* to update peer list contents. For quota redistribution protocol, we should have APIs such as *JoinRequest (Instance I)* to request to join the current DCaaS peer network, *UpdateAck (Instance I)* to inform leader with updates confirmation, and

*joinAck(DCP P, PeerList R)* to acknowledge the acceptance of a new DCaaS instance. For stabilization Protocol, we should have APIs such as *StabReq(Object x)* to request stabilization of a given object, *StabRes(Object x)* to respond with the object value, and *StabCom(Object x)* to enforce a common object value, and finally *StabAck(Object x)* to acknowledge object stabilization. For quota borrowing protocol, we should have APIs such as *QBrwReq(Object x, Amount y)* to request borrowing a certain amount of quota, *QuotaTransfer(Object x, Amount y)* to transfer certain amount of quota to another DCaaS instance, and *QuotaTrAck(Object x, Amount y)* to acknowledge the quota transfer process. For leader election protocol, we should have APIs such as *LeaderReq(Instance I)* to nominate a leader, *LeaderAck(DCP P, PeerList R)* to accept a leader nomination, *Synch(DCP P, PeerList R)* to synchronize DCaaS instances, and *SynchAck* to acknowledge the success of the synchronization process. Of course all DCaaS APIs will be under proper security management; however security is out of the scope of this paper.

### C. DCaaS Recovery

In case of a given DCaaS instance failure. We will use classical DB recovery approaches using data logs for recovering eventual consistency objects to rollback any uncommitted transactions, while for session consistency objects, we will just fetch the last values from DB. The problem will be in the strong consistency objects, as the allocated quota for strong consistency objects has to be redistributed among remaining DCaaS instances. Quota redistribution is done when the leader or any other DCaaS instances noticed the failure of such DCaaS instance. Hence, it sends *UpdatePeerList* request to all the DCaaS instances in the peer list, so they can remove such instance from their peer list and update their SCOQ list. When a DCaaS instance recovers from failure, it follows the join protocol in Figure 5 to rejoin the DCaaS peer network. It is important to note that join request is idempotent, hence if multiple copies of the same join request are somehow created, they will have the same effect and no problems could occur. It is also important to note that when a DCaaS instance receives a request for adding a new instance, it checks its log to see if it has a previous history with this instance that if there exist any unfinished communications or acknowledgments so that they can pursue it. The recovery problem becomes more complicated in case of a leader failure during a given protocol execution. In this case, DCaaS instances who still alive could need to elect a different leader to accomplish the required tasks. For example, in case of join protocol, DCaaS instances will send their update acknowledgments to the new instance directly if they notice leader failure. In this case, the new instance will receive multiple join acknowledgments, which will not cause a problem as the join acknowledgments operation is idempotent. However, if the new DCaaS instance times out for not receiving any join acknowledgement, it could resubmit its request to another DCaaS instance. In case of the leader failure during stabilization protocol (see Figure 3), if it failed before receiving stabilization responses, we will have no problems as no DCPs have changed, however if it failed before receiving all stabilization acknowledgments this means we could have a problem. As some DCaaS instances could have successfully received the stabilization command and updated their DCPs while other instances could not do such updates, if the leader recovers back it could pursue the stabilization process with the remaining DCaaS instances, however if the leader could not recover, this means we have a DCP inconsistency problem as different DCaaS instances will have different versions of the DCP. This problem will be solved after the election of a new leader that will make sure all DCaaS instances are using a common DCP configuration. Leader election occurs when one or more of the follower instances notice the leader failure, and broadcast the election request. Leader election process occurs as depicted in Figure 6. First, a DCaaS instance broadcasts to other instances a request for being the leader, other instances could accept and respond by their (compressed) DCPs and peer lists or reject the request. If majority of instances accepted, this means a new leader is elected, otherwise elections has to be repeated. Once a new leader is elected, the new leader starts synchronizing other DCaaS instances to have a common DCP and peer list. Once a DCaaS instance receives a synchronization request, it updates its information and computes its new SCOQ. Of course we could adopt different strategies for choosing a common DCP and a common peer list. The simplest strategy is choosing the most recent ones. Other strategies is to choose the most restrictive ones, the least restrictive ones, ones leading to least cost updates, etc.

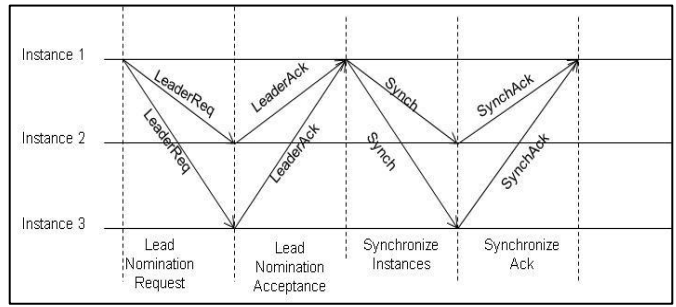

Fig. 6. Stages of Leader Election Protocol.

In our implementation, we let the choice of the DCP selection strategy as one of the configuration parameters for DCaaS services. Of course, there is no optimal strategy as each one has pros and cons. Comparison between strategies is out of the scope of this paper due to space limitation. In case of the leader failure during the quota borrowing protocol (see Figure 2), we will have no problems if failure occurs before receiving the *QuotaTransfer* messages as no quota has actually transferred. However, if failure occurs after sending *QuotaTrack* messages, this means transferred quotas are lost. Again, if the leader recovered before an election request generation, the quotas will be recovered; otherwise quotas will be redistributed during the new leader election process. It is important to note that there is impossibility of distributed consensus with one faulty process [10]. Hence, fault tolerance algorithms should be based on majority rather than complete consensus. Therefore, implementations of proposed approaches adopt a majority of instances (as in Paxos algorithms [17-19]), where f is the number of expected failed instances. We are also extending this approach to handle SaaS requests in order to ensure Byzantine fault tolerance [14] for SaaS services. This is achieved by allowing the a SaaS service to send its requests to multiple DCaaS service instances, and then accept a response only when it is returned by majority of DCaaS instances, details of this approach is out of the scope of this paper.

## VII. EXPERIMENTS

We performed basic simulation experiments using the cloudsim tool [25] that enables us to simulate cloud environments. For simplicity, we assumed that we have only one DCaaS instance per cloudlet; we have two identical cloudlets (i.e. datacenters) with WAN connection of 500ms, and one user-base accessing the first cloudlet with latency 50ms. This user-base generates 1000 request per hour, and each request contains one read and one write operations for a strong consistency data object. We simulated both cloudlets with 5 virtual machines each. Each virtual machine contains 512 MB and 1KB bandwidth. Each cloudlet is build using two 4-core processors identical servers with 10000 MIPS, 200GB RAM, 10 TB storage, and IMB bandwidth. We run the simulation for period of 1 day and computed the average, minimum and maximum response time for the whole user-base. In our experiments, we compare between the pure locking approach (that locks record on both cloudlets for every request), against the proposed DCaaS service approach. We assumed all objects in the DCP require strong data consistency. However, to show the borrowing effect, we repeated the experiment, by adopting different global quota borrowing rates from outside the cloudlet, which are 0%, 10%, and 50%, which means are 0%, 10%, and 50% of the data accesses will require quota borrow operation, respectively. We choose to compare global quota borrowing (among cloudlets) rather than local quota borrowing as global quota borrowing is the main process that could negatively affect response time due to access of WAN connections. However, local quota borrowing process occurs internally inside the cloudlet where latency is small, hence it will not have a huge impact of response time. Experiments results are listed in Table 1.

| Approach | Avg (ms) | Min (ms) | Max (ms) |
|---|---|---|---|
| Locking Approach | 1600 | 1038 | 2242 |
| DCaaS with 0% Quota Borrow | 200 | 150 | 258 |
| DCaaS with 10% Quota Borrow | 350 | 249 | 465 |
| DCaaS with 50% Quota Borrow | 600 | 414 | 815 |

As we can see, when the quota is enough (i.e. 0% quota borrow), the DCaaS instance does not need to communicate with the other DCaaS instance through the WAN, as all requests are fulfilled within the cloudlet; hence response time is drastically improved (i.e. from 1600ms to 200 ms). However, when DCaaS needs to borrow quota from the other cloudlet response time starts to increase as WAN connection is used, for example when 50% quota borrow is required response time becomes 3 times worse (i.e. 600 ms). Based on results in Table 1, we conclude that response time increases when the global quota borrowing percentage increases. Hence, to minimize such response time, we argue that the initial quota distribution among cloudlets should be based on their user-based demand rates, that cloudlet with higher demand should get higher percentages of the quota.

To show the effect of DCP adoption on performance, we conducted a similar experiment when the percentage of strong data objects in the DCP is 0%, 10%, 50%, and 100% respectively. We assumed that the global quota borrow percentage is 50% to have comparable results with Table 1. Experiments results are listed in Table 2. As we can see, when we have no strong data objects (i.e. the 0% DCP case) the response time improves as no need for borrowing operations at all. We achieved much better performance (i.e. 50 ms) when compared with the 0% global quota borrow case in Table 1 (i.e. 200 ms). This is because there is no locking is required to maintain local correctness. However, when we started to increase the percentage of the strong data objects in the DCP, response time starts to increase as quota borrowing operations are required, which require access for WAN connections. Hence, we conclude that to improve performance, we should minimize the percentage of the strong data objects in the DCP. However, in case of having strong data objects in the DCP, the initial quota distribution between cloudlets should be distributed in a manner that minimizes the global quota borrow rate. We argue that the quota should be distributed according to the cloudlet user-base demand rate.

| Approach | Avg (ms) | Min (ms) | Max (ms) |
|---|---|---|---|
| DCP with 0% Strong Consistency Objects | 50 | 36 | 65 |
| DCP with 10% Strong Consistency Objects | 66 | 48 | 129 |
| DCP with 50% Strong Consistency Objects | 200 | 141 | 260 |
| DCP with 100% Strong Consistency Objects | 600 | 414 | 815 |

## VIII. CONCLUSION AND FUTURE WORK

In this paper, we argued that strong consistency requirements should be adopted only for data objects crucial for application correctness, otherwise weaker forms of data consistency should be adopted. Therefore, we proposed to use the concept of data consistency plan (DCP) to define the consistency requirements for SaaS services, and proposed to use a new platform service (i.e. Data Consistency as a Service (DCaaS)) for executing such DCP plan to decouple SaaS developers from managing data uncertainty issues in their code. We also proposed a quota-based approach for managing data uncertainty on eventually consistent cloud data stores. The proposed approach ensures global data consistency by distributing the capacity of strong consistency data objects among datacenters, and then adopts a lazy replication approach for synchronizing the data stores. Experiments show that proposed quota-based approach realized by the DCaaS service provides much better response time when compared with locking and blocking techniques.

Future work will be mainly focused on providing SaaS developers more flexibility for defining the service DCP, by allowing an object to have different consistency levels at the same time; depending on the performed SaaS operations and customers' SLAs. This will be achieved by having a new object model that adopts different uncertainty modeling and analysis techniques.

### REFERENCES

[1] D. Barbar´a, H. Garcia-Molina, and D. Porter. The management of probabilistic data. IEEE TKDE, 4(5), 1992.

[2] O. Benjelloun, A.D. Sarma, A. Halevy, and J. Widom. ULDB: Databases with Uncertainty and Lineage. In VLDB, 2006.

[3] R. Buyya, J. Broberg, and A. Goscinski (eds), "Cloud Computing: Principles and Paradigms", ISBN-13: 978-0470887998, Wiley Press, New York, USA, March 2011.

[4] F. Chang, J. Dean, S. Ghemawat, W. C. Hsieh, D. A. Wallach, M. Burrows, T. Chandra, A. Fikes, and R. E. Gruber. Bigtable: A Distributed Storage System for Structured Data. In OSDI, pages 205–218, 2006.

[5] B. F. Cooper, R. Ramakrishnan, U. Srivastava, A. Silberstein, P. Bohannon, H.-A. Jacobsen, N. Puz, D. Weaver, and R. Yerneni. PNUTS: Yahoo!'s hosted data serving platform. Proc. VLDB Endow., 1(2):1277–1288, 2008.

[6] S. Das, D. Agrawal, and A. E. Abbadi. G-Store: A Scalable Data Store for Transactional Multi-Key Access in the Cloud. In SoCC, 2010.

[7] S.B. Davidson, H. Garcia-Molina, and D. Skeen, "Consistency in partitioned networks", ACM Comput. Surv, vol. 17, no. 3, pp.341–370, 1985.

[8] G. DeCandia, D. Hastorun, M. Jampani, G. Kakulapati, A. Lakshman, A. Pilchin, S. Sivasubramanian, P. Vosshall, and W. Vogels. Dynamo: Amazon's highly available key-value store. In SOSP, pages 205–220, 2007.

[9] A. Demers, D. Greene, C. Hauser, W. Irish, J. Larson, S. Shenker, H. Sturgis, D. Swinehart, and D. Terry, "Epidemic algorithms for replicated database maintenance", in Proc. of ACM Conference on Principles of Distributed Computing (PODC'87), 1987.

[10] M. Fischer, N. Lynch, and M. Paterson. Impossibility of Distributed Consensus With One Faulty Process. Journal of the ACM, 32(2), 1985.

[11] S. Gilbert and N. Lynch, "Brewer's Conjecture and the Feasibility of Consistent, Available, Partition-Tolerant Web Services", SIGACT News, vol. 33, no. 2, 2002.

[12] J. Gray, P. Helland, P. O'Neil, and D. Shasha, "The dangers of replication and a solution", in Proc. of ACMSIGMOD International Conference on Management of Data, pp. 173–182,1996.

[13] P. Helland. Life beyond distributed transactions: an apostate's opinion. In CIDR, pages 132–141, 2007.

[14] R. Kotla, L. Alvisi, M. Dahlin, A. Clement, and E. Wong," Zyzzyva: Speculative byzantine fault tolerance", In Symposium on Operating Systems Principles (SOSP), 2007.

[15] T. Kraska, M. Hentschel, G. Alonso, and D. Kossmann, "Consistency Rationing in the Cloud: Pay only when it matters", in Proc. of the international Conference on VLDB, 2009.

[16] L. Lamport. Time, Clocks, and the Ordering of Events in a Distributed System. Commun. ACM, 21(7), 1978.

[17] L. Lamport. The part-time parliament. ACM Transactions on Computer Systems, 16(2):133–169, May 1998.

[18] L. Lamport. Fast Paxos. Distributed Computing, 19(2):79–103, Oct. 2006

[19] L. Lamport. Lower bounds for asynchronous consensus. Distributed Computing, 19(2):104–125, Oct.2006.

[20] D. B. Lomet, A. Fekete, G. Weikum, and M. J. Zwilling. Unbundling transaction services in the cloud. In CIDR Perspectives, 2009.

[21] K. Manassiev and C. Amza, "Scalable database replication through dynamic multiversioning",in Proc. Centre for Advanced Studies on Collaborative research, 2005.

[22] H. Wada, A. Fekete, L. Zhao, K. Lee and A. Liu, "Data Consistency Properties and the Trade-offs in Commercial Cloud Storages: the Consumers' Perspective", in Proc. of the 5th biennial Conference on Innovative Data Systems Research, 2011.

[23] W. Vogels, "Eventually Consistent", ACM Queue vol. 6, no. 6, December, 2008.

[24] Cassandra, available: http://cassandra.apache.org/

[25] CloudSim, available : http://www.cloudbus.org/cloudsim/.

[26] I. Elgedawy, B. Srivastava, and S.Mital, "Exploring Queriability of Encrypted and Compressed XML Data", In proceedings of the 24th of the International Symposium on Computer and Information Sciences, Northern Cyprus, 2009.

[27] I. Elgedawy, "Data Consistency as a Service (DCaaS)", submitted to the 27th of the International Symposium on Computer and Information Sciences, France, 2012. Submission number 11.

[28] I. Elgedawy, "On-demand conversation customization for services in large smart environments", IBM Journal of Research and Development, Special issue on Smart Cities, Vol. 55, No. 1/2, January 2011.